# Correlation-driven insulator-metal transition in near-ideal vanadium dioxide films

A. X. Gray[1,2★], J. Jeong[3], N. P. Aetukuri[3], P. Granitzka[1,4], Z. Chen[1,5], R. Kukreja[1,6], D. Higley[1,7], T. Chase[1,7], A. H. Reid[1], H. Ohldag[8], M. A. Marcus[9], A. Scholl[9], A. T. Young[9], A. Doran[9], C. A. Jenkins[9], P. Shafer[9], E. Arenholz[9], M. G. Samant[3], S. S. P. Parkin[3] and H. A. Dürr[1★]

[1]Stanford Institute for Materials and Energy Sciences, SLAC National Accelerator Laboratory, 2575 Sand Hill Road, Menlo Park, California 94025, USA

[2]Department of Physics, Temple University, 1925 N. 12th St., Philadelphia, Pennsylvania 19130, USA

[3]IBM Almaden Research Center, 650 Harry Road, San Jose, California 95120, USA

[4]Van der Waals-Zeeman Institute, University of Amsterdam, 1018XE Amsterdam, The Netherlands

[5]Department of Physics, Stanford University, Stanford, California 94305, USA

[6]Department of Materials Science and Engineering, Stanford University, Stanford, California 94305, USA

[7]Department of Applied Physics, Stanford University, Stanford, California 94305

[8]Stanford Synchrotron Radiation Lightsource, SLAC National Accelerator Laboratory, 2575 Sand Hill Road, Menlo Park, California 94025, USA

[9]Advanced Light Source, Lawrence Berkeley National Laboratory, One Cyclotron Road, Berkeley, California 94720, USA

★email: axgray@temple.edu, hdurr@slac.stanford.edu

We use polarization- and temperature-dependent x-ray absorption spectroscopy, in combination with photoelectron microscopy, x-ray diffraction and electronic transport measurements, to study the driving force behind the insulator-metal transition in $VO_2$. We show that both the collapse of the insulating gap and the concomitant change in crystal symmetry in homogeneously strained single-crystalline $VO_2$ films are preceded by the purely-electronic softening of Coulomb correlations within V-V singlet dimers. This process starts 7 K (±0.3 K) below the transition temperature, as conventionally defined by electronic transport and x-ray diffraction measurements, and sets the energy scale for driving the near-room-temperature insulator-metal transition in this technologically-promising material.

A clear understanding of how new electronic and structural phases of matter arise and evolve is not only important for basic science but is also becoming crucial for information technology. Our ability to navigate and ultimately control the complex dynamical pathways in the multidimensional landscape of the electronic, spin, and lattice degrees of freedom is starting to play a critical role in achieving technical feasibility and efficient performance of future electronic devices [1].

One of the prime material candidates for such devices, vanadium dioxide ($VO_2$), undergoes an insulator-metal transition with a four-orders-of-magnitude increase in conductivity and a concomitant monoclinic-rutile structural transformation accompanied by the dimerization of neighboring vanadium atoms just above room temperature [2]. This makes $VO_2$ suitable for technological applications, from solid-state sensors and optical detectors to field-effect transistors and memristors [1], and motivates investigations aimed at controlling the insulator-metal transition by external stimuli [3-5] and epitaxial strain [6].

The nature of the driving force behind the insulator-metal transition in this prototypical strongly-correlated electron system is one of the longest-standing problems in condensed matter physics [2]. Both electron-electron correlations and electron-lattice interactions are believed to be relevant. However, the question of which drives the other and the interplay between Mott-Hubbard [7,8] and Peierls [9,10] mechanisms remain under debate [11-15].

The structural and electronic dichotomy of the insulator-metal transition in $VO_2$, depicted schematically in Fig. 1, opens the door for two possible explanations as to what could be the driving force and the physical mechanism behind this phenomenon. On one hand, a structural transformation with strong dimerization (Figs. 1(a),(c)) presents seemingly convincing evidence for the Peierls-like picture, wherein opening of the insulating gap shown schematically in Fig. 1(d) is caused by the lattice distortion [9,10]. On the other hand, the evidence of strong electron-electron correlation effects suggests the Mott-Hubbard scenario in which strong Coulomb interaction between electrons plays the key role in triggering the insulator-metal transition by splitting the near-Fermi-level electronic states (Fig. 1(b)) into the bonding and anti-bonding bands, thus opening-up an insulating gap (Fig. 1(d)) [7,8]. To this day, however, the underlying physics of the transition remains elusive due to the lack of success in consolidating all of its experimentally observed structural and electronic aspects in a singular self-consistent quantitative theoretical picture [7-14]. A further complication arises due to phase separation scenarios often accompanying insulator-metal transitions [16-18]. In bulk-like $VO_2$ films the coexistence of insulating and metallic-like patches was observed in various experiments probing electronic and structural aspects [16,17]. Similarly, in thick epitaxial films the formation of unidirectional metallic-like stripes [18] and a monoclinic-like metallic phase [19] was observed in the vicinity of the transition.

Here we show that near-ideal, high-quality ultrathin epitaxial $VO_2$ films grown on $TiO_2$(001) substrates display a spatially homogeneous insulator-metal transition without phase segregation. We track the $VO_2$ insulator-metal transition by monitoring with polarization-dependent x-ray absorption spectroscopy (XAS) spectroscopic features of the dimer V-V electronic correlations (Fig. 2d), the electronic band gap (Fig. 2c) and the dimer V-V Peierls lattice distortion (Fig. 2d), respectively. This enables the unambiguous assignment of the insulator-metal transition to two consecutive processes occurring with increasing temperature. Initially, the insulating phase is modified by a weakening of electronic correlations in dimer V-V singlet states [15]. This process sets in as much as 7 K (±0.3 K) below the proper insulator-metal transition temperature ($T_{IMT}$ = 295±0.3 K) [20] probed by the electronic band gap collapse. The gap collapse starts as soon as the V-V singlet correlations have completely disappeared at $T_{IMT}$. Only at $T_{IMT}$ do we detect structural changes in V-V dimerization [20] as implied by a Peierls scenario, as well as electronic changes consistent with the disappearance of a dimer V-V Peierls lattice distortion.

For our experiment, high-quality single-crystalline $VO_2$ thin films were grown epitaxially on $TiO_2$(001) substrates by pulsed laser deposition, following the procedures described in-depth in Refs. 5 and 6 (our prior studies). This results in coherently-strained 10 nm-thick $VO_2$(001) films with a $c_R/a_R$ lattice-constant ratio of 0.617 [20,6]. The films undergo an abrupt insulator-metal transition at 295 K upon heating that is accompanied by a change in crystal symmetry. This transition is sharper than that observed for comparable films [19] and indicative of the absence of any detectable electronic phase separation (see below).

Temperature- and x-ray polarization-dependent XAS measurements were carried out at the elliptically polarized undulator beamline 4.0.2 of the Advanced Light Source, using the *Vector Magnet* endstation equipped with a 4-axis sample manipulator with cryogenic cooling as well as conductive heating capabilities [21]. The nearly 100% linearly-polarized x-ray beam was focused down to 100 μm diameter spot at the sample surface. Average probing depth in the total electron yield XAS detection mode was estimated to be about 5 nm, providing bulk-sensitive information with minimal contribution from surface adsorbates. The temperature was varied using a closed-loop feedback-activated temperature controller, providing temperature stability of ±0.1 K. Measurements were carried out at several locations on the sample, to exclude the possibility of x-ray sample damage.

Figure 2 shows high-resolution polarization-dependent XAS measurements at the O *K* edge. O *K* edge XAS probes the O 2*p*-projected unoccupied density of states resulting from dipole-allowed x-ray transitions from the 1*s* core shell. It is well known [24-26] that this provides access to all relevant states in the unoccupied $VO_2$ conduction band via O 2*p* - V *3d* orbital hybridization. We use the x-ray polarization dependence of XAS to determine the orbital symmetry. In the grazing incidence experimental geometry shown in Fig. 2(a), linearly polarized x-rays with the electric-field orientation parallel to the $c_R$ axis ($E \parallel c_R$) preferentially probe the unoccupied $d_{\parallel}$ states of $a_{1g}$ symmetry. $d_{\parallel}$ orbitals are preferentially oriented along the $c_R$ direction (see Fig. 2(b)). X-rays with their electric-field orientation perpendicular to the $c_R$ axis ($E \perp c_R$) are more sensitive in probing the states with $e_g^{\pi}$ symmetry, comprised of $d_{xz}$ and $d_{yx}$ orbitals ($\pi$* states) (see Fig. 2(b)) [6, 24].

Figure 2(c) shows typical temperature and polarization-dependent O XAS spectra for the $VO_2$ film, as measured for the photon energy range between 528 and 531 eV. The spectra have been normalized to equal edge jump accounting for the incomplete alignment of x-ray polarization with the $d_{\parallel}$ orbital orientation and the varying orbital multiplicity for the two polarizations [24,27]. The leading edge found at 528.7 eV (in the insulating phase) is known to sensitively probe the closing of the electronic band gap with increasing temperature [24]. This is observed as the edge shift to lower photon energy with increasing temperature which is one of the major effects seen in Fig. 2(c) especially between spectra taken with fixed x-ray polarization. Overlaid with this change in band gap are polarization effects seen in the insulating phase but not for metallic $VO_2$. Two most prominent polarization-dependent differences are observed at the photon energies of 528.7 eV and 530.2 eV (see also Fig. 2(d)) and are related to the dimer V-V electronic correlations and Peierls lattice distortion, respectively. We will first describe these three spectroscopic features in Figs. 2(c) and 2(d) before proceeding to the temperature dependence.

The XAS spectra measured at the temperature of 310 K, well above the insulator-metal transition ($T_{IMT}$=295 K), are shown in Fig. 2(c) as open and solid red symbols. These spectra are good representations of the metallic unoccupied density of states immediately above the Fermi level [6,28]. Consistent with prior experimental results [24] and theoretical calculations [15] no significant difference between the spectra collected using two mutually-orthogonal x-ray linear polarizations is observed due to an almost isotropic distribution of the $t_{2g}$ orbitals near the Fermi level [15,27].

The insulating-state spectra (shown in blue) collected at 280 K (15 K below the insulator-metal transition) are shifted by about 80-100 meV to higher photon energy due to the opening of the insulating gap and, in stark contrast with the metallic-state spectra, are significantly different. Their intensity difference plotted in Fig. 2(d) features two major polarization-dependent peaks favoring $E \parallel c_R$ polarization. The more prominent peak centered at the photon energy of 530.2 eV is a well-known $d_\parallel$ state arising from the Peierls distortion of the lattice [11,15,24]. This $d_\parallel$ *Peierls* peak is depicted schematically (red outline) in Fig. 1(d). The second feature is observed at the photon energy of 528.7 eV. Such additional XAS intensity for $E \parallel c_R$ presents direct evidence of the additional $d_\parallel$ orbital character at the onset of the conduction band (depicted schematically in Fig. 1(d)) which has been identified as the fingerprint of the strong electronic correlations within the V-V dimers [15,24]. In addition to being present in the spectra for the coherently-strained thin $VO_2$(001) films, such a $d_\parallel$ V-V singlet state has also been observed for the bulk $VO_2$ single-crystals [24]. It represents a unique feature of the strongly-correlated insulating phase of crystalline $VO_2$.

The temperature dependence of the electronic band gap (red solid symbols in Fig. 3(a)) displays the insulator metal transition with $T_{IMT}$ = 295±0.3 K. This insulator-metal transition is closely tracked by the monoclinic-rutile structural transition [20] monitored by the dimer V-V Peierls peak (yellow solid symbols in Fig. 3(a)). Extended datasets with individual spectra for each temperature are presented in Fig. S2 of [20]. Interestingly, the temperature dependence of the dimer V-V singlet state is very different (blue/white symbols in Fig. 3(a)). We observe that the V-V state intensity (as defined in Fig. 2(d)) begins to decay at 288 K which is 7 K below $T_{IMT}$. The dimer V-V singlet state disappears completely just before the onset of the electronic band gap collapse at 294 K. Thus, our results strongly suggest that the insulator-metal transition in $VO_2$ follows a three-stage pathway depicted schematically in Figs. 3(b)-(d). **(1)** At temperatures that are up to 7 K below the insulator-metal transition (denoted $T \ll T_{IMT}$ in Fig. 3(a)), $VO_2$ is in the insulating monoclinic phase, with two $3d^1$ electrons of two adjacent dimer V-V atoms forming a strongly-correlated singlet state of $d_\parallel$ symmetry [15] (Fig. 3(b)). **(2)** Upon heating above $T_{IMT}$-7K, $VO_2$ remains in the insulating monoclinic state, i.e. the electronic band gap remains unchanged as demonstrated by the upper band gap edge shown in Fig. 3(a). However, the electronic correlations start to soften, as evidenced by the decay of the dimer V-V singlet state intensity (Fig. 3(c)). **(3)** Finally, once the electronic correlations are sufficiently diminished, the band gap collapse is initiated. This latter process is accompanied by the change in crystal symmetry from insulating monoclinic to metallic rutile as evidenced by the dimer V-V Peierls peak change (Fig. 3(d)). Above 297 K $VO_2$ is in a homogeneous metallic state. Process **(3)** seems to correspond to the conventional Peierls mechanism where lattice distortions keep the band gap intact. However, our results show that it is the preceding decay of electron correlations giving rise to dimer V-V singlet state that set the high-temperature energy scale in $VO_2$. As an important consequence of this transition pathway, one can define a second distinct critical temperature ($T_{corr}$=290 K ± 0.3 K) at which $VO_2$ undergoes a purely-electronic transition between a strongly-correlated and a conventional monoclinic Peierls insulator. This transition temperature $T_{corr}$ is 5 K (± 0.3 K) below the insulator-metal transition.

We finally need to address the question of spatial homogeneity of the observed phenomena since phase segregation scenarios have been observed for bulk-like and thin-film $VO_2$ [16-19]. A temperature- and polarization-dependent spectro-microscopic investigation of the sample using photoemission microscopy (PEEM) was carried out at the EPU beamline 11.0.1.1 of the Advanced Light Source, using the *PEEM-3* microscope routinely facilitating sub-50 nm spatial resolution. Measurements at the O *K* absorption edge (528.7 eV) reveal that our thin coherently-strained $VO_2$ film grown in (001) crystallographic orientation on a single-crystalline $TiO_2$ substrate does not undergo any detectable insulator/metal phase segregation across the insulator-metal transition (see Fig. 3(e)). We note that the all-over PEEM contrast in the images of Fig. 3(e) follows the electronic band gap collapse in Fig. 3(a) when normalized to defects (dark dots in Fig. 3(e)). Indirectly, such quasi-instantaneous single-domain switching is evidenced by the sharpness of the transition (2-4 K width) observed in the electronic structure via XAS and electronic transport measurements, as well as structurally via x-ray diffraction spectroscopy [20]. This is in stark contrast to bulk-like films where epitaxial strain is fully relaxed. We find that such films do display a separation into metallic and insulating regions during the insulator-metal transition as shown in Fig S3 of [20].

In summary, our results indicate that the temperature-driven insulator-metal transition in a prototypical strongly-correlated oxide $VO_2$ is preceded and, possibly, driven by the purely-electronic phase transition occurring at a lower temperature ($T_{corr}<T_{IMT}$). We have observed the emergence and the temperature-dependent evolution of the new intermediate monoclinic/insulating phase in $VO_2$ which is characterized by softening and disappearance of the strong correlations within the V-V dimers. These findings have a far-reaching impact on our understanding of the complex physics of the insulator-metal transition in strongly-correlated oxides, since they suggest that the fundamental electronic and structural transformations in these materials arise from *precursory* changes in the electronic correlations. Thus, for all the future practical applications of $VO_2$ as a building-block for next-generation electronic devices, as well as for a wide range of time-resolved pump-probe experiments, this underpinning transition and the characteristic $T_{corr}$ should be carefully characterized and understood.

Research at Stanford was supported through the Stanford Institute for Materials and Energy Sciences (SIMES) under contract DE-AC02-76SF00515 and the LCLS by the US Department of Energy, Office of Basic Energy Sciences. The Advanced Light Source is supported by the Director, Office of Science, Office of Basic Energy Sciences, US Department of Energy under Contract No. DE-AC02-05CH11231. Authors would like to thank C.-C Chen, B. Moritz, T. P. Devereaux and M. van Veenendaal for helpful discussions.

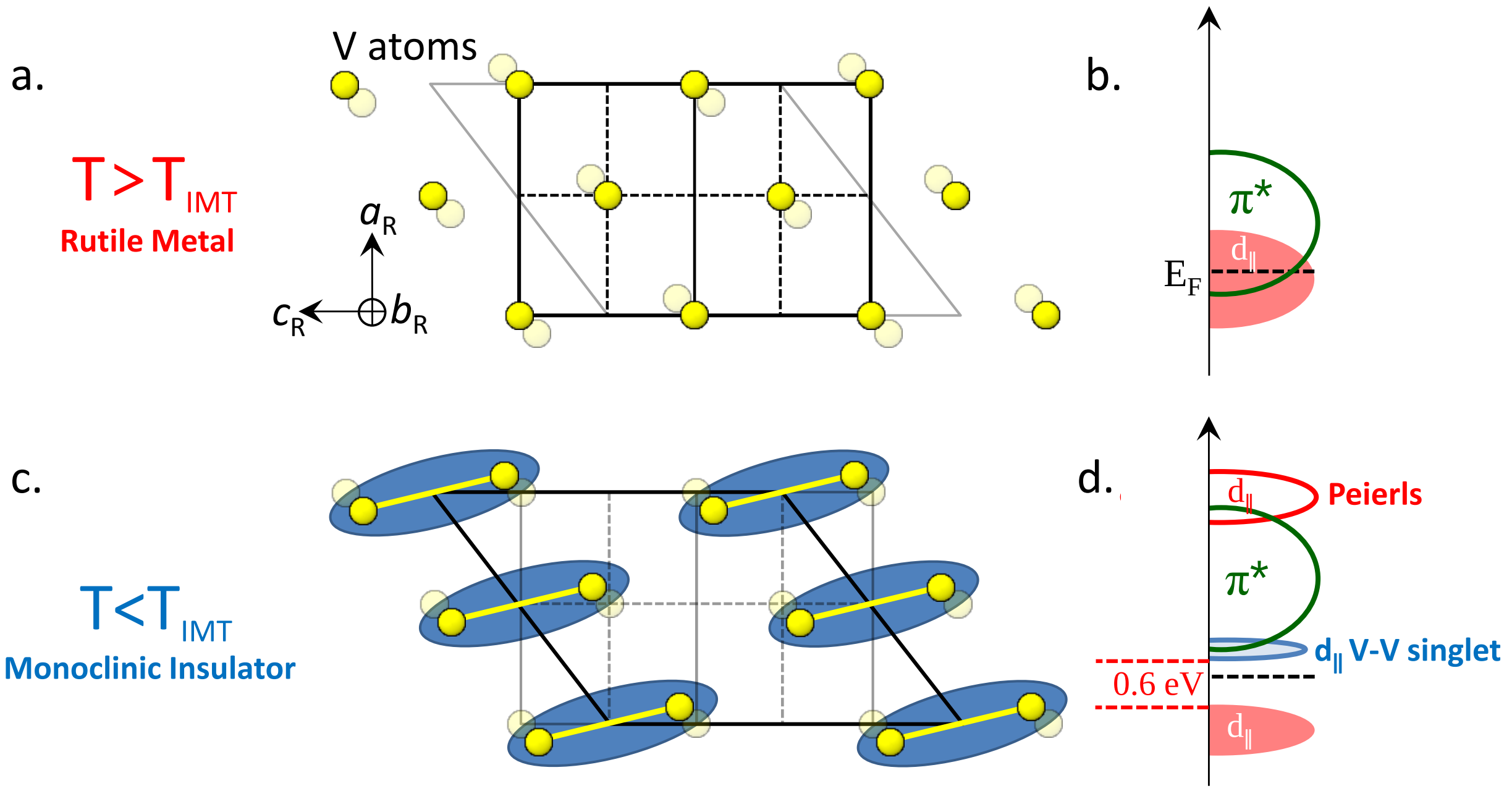

**FIG. 1.** **(a)** In the high-temperature (T>T$_{IMT}$) metallic phase $VO_2$ forms rutile lattice structure with $P4_2/mnm$ space symmetry where V atoms (yellow) occupy centers of six-fold oxygen-coordinated sites (O atoms not shown). **(b)** The near-Fermi-level $t_{2g}$ states are separated in energy by the orthorhombic component of the crystal field into the twofold-degenerate $e_g^{\pi}$ ($\pi$*) states and a single $a_{1g}$ orbital, which is aligned parallel to the rutile $c$ axis ($c_R$) and is thus commonly denoted $d_\parallel$. The two bands overlap in energy, and the resulting non-zero density of states at the Fermi level accounts for the metallic behavior of the rutile phase [11,15]. **(c)** In the low-temperature (T<T$_{IMT}$) insulating phase the lattice undergoes a structural transition to a lower-symmetry ($P2_1/c$) monoclinic crystal system via dimerization of the neighboring V atoms along the $c_R$ direction and tilting of the resultant V-V dimers along the rutile [110] and [1$\bar{1}$0] directions. Each dimerized V-V atomic pair shares a singlet electronic state composed of two strongly-correlated V 3$d^1$ electrons (shown in blue) [15]. **(d)** V-V dimerization splits the highly-directional $d_\parallel$ orbitals into the bonding and anti-bonding bands, and the tilting of the dimers shifts the $\pi$* band to higher energies due to the increase of the $p$-$d$ orbital overlap [11,15], together producing an insulating gap of 0.6 eV – the key aspect of the monoclinic phase observed via a wide variety of experimental techniques [22,23]. Additional $d_\parallel$ feature in DOS, which arises at the onset of the conduction band (shown in blue) is a unique fingerprint of the strong electron-electron correlations within the dimers [15], accessible experimentally via polarization-dependent XAS at the O $K$ edge (see Ref. 24 and this work).

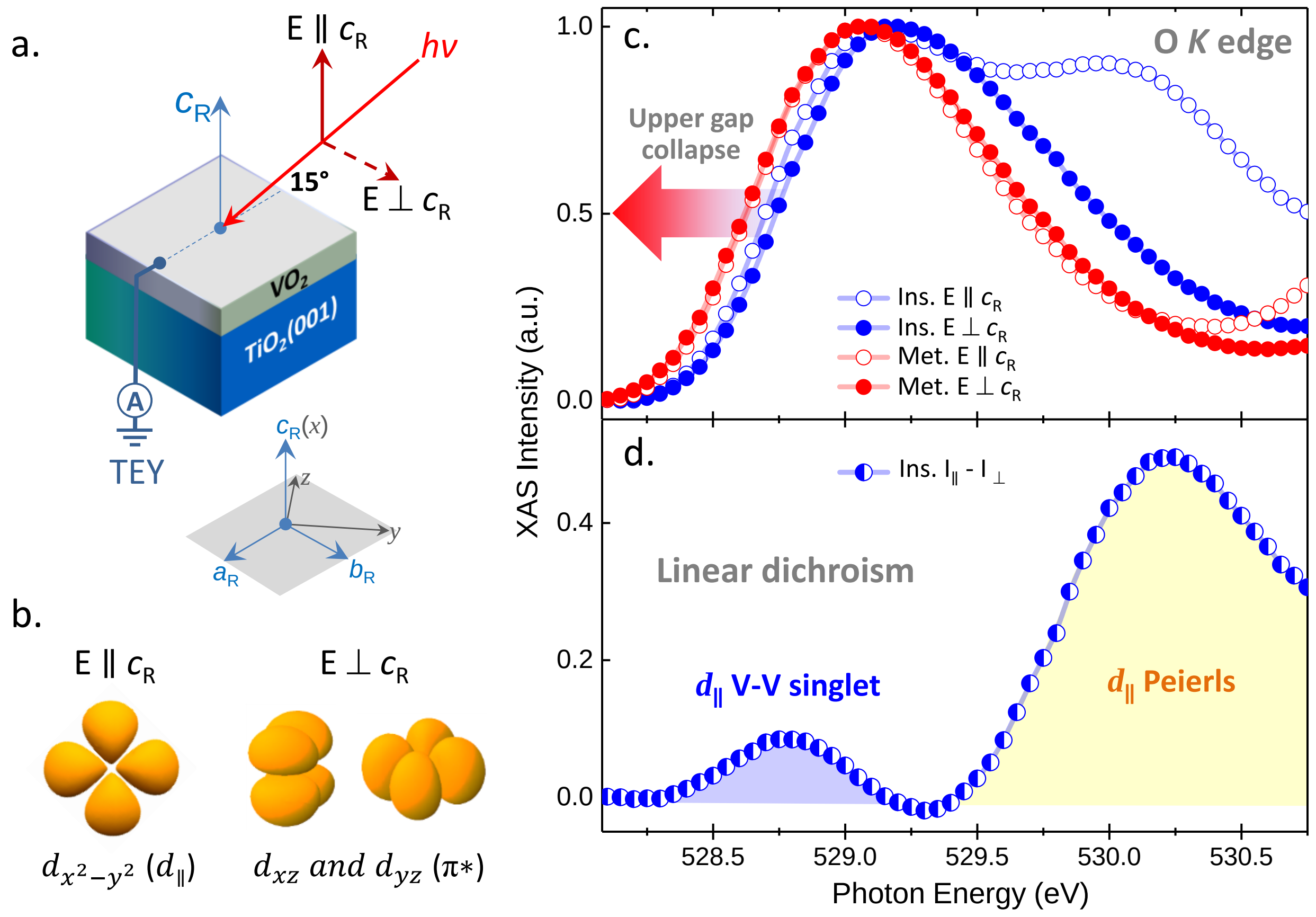

**FIG. 2.** **(a)** Schematic diagram of the polarization-dependent XAS measurement geometry used in this study. Linearly-polarized x-rays are incident at 15° to the surface of the coherently-strained epitaxial 10 nm-thick $VO_2/TiO_2(001)$ sample, with the photon polarization set to either parallel (E ∥ $c_R$) or perpendicular (E ⊥ $c_R$) to the $c_R$ axis (along the sample normal) of the film enabling preferential probing of the strongly-directional $d_\parallel$ ($d_{x^2-y^2}$) and $\pi^*$ ($d_{xz}$ and $d_{yx}$) orbitals depicted in **(b)**, respectively. **(c)** Polarization-dependent O *K* edge XAS measurements of $VO_2$ in the high-temperature ($T=T_{IMT}+15K$) metallic state (shown as open and solid red symbols) and in the low-temperature ($T=T_{IMT}-15K$) insulating state (shown as open and solid blue symbols). Shift of the leading slope of the edge towards lower energy between the insulating and the metallic phases of $VO_2$ corresponds to the collapse of the insulating band gap on the unoccupied side of the energy band diagram (upper gap collapse). Distinct dichroic signal at the onset of the absorption edge in the insulating state presents direct evidence of the additional $d_\parallel$ orbital character at the bottom of the unoccupied conduction band, which is predicted by theory [15] and is shown schematically in Fig. 1(d). **(d)** $d_\parallel$ V-V singlet peak at the onset of the conduction band (528.7 eV) and the $d_\parallel$ Peierls peak at 530.2 eV, both obtained by calculating the $I_\parallel$-$I_\perp$ XAS intensity taken directly from **(c)**.

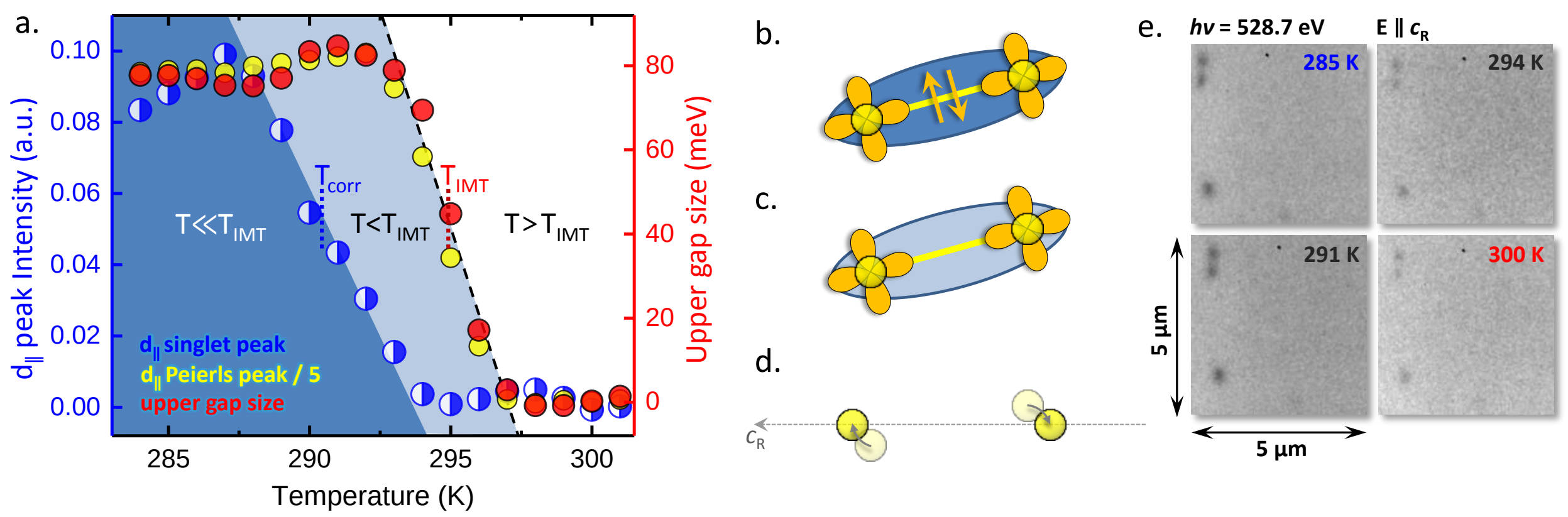

**FIG. 3.** **(a)** Temperature-dependent evolution of the $d_{\parallel}$ V-V singlet peak intensity (blue/white symbols) showing distinctly different transition temperature ($T_{corr}$=290 K) as compared to the $T_{IMT}$ (295 K), the critical temperature at which the collapse of the upper band gap (red solid symbols) as well as the structural transition (yellow solid symbols) is observed in $VO_2$. **(b)-(d)** Schematic representation of the three phases of $VO_2$ implied by the experimental data in the plots on the left. **(b)** The low-temperature phase ($T \ll T_{IMT}$) is a monoclinic insulator with a strongly-correlated singlet electronic state on each V-V dimer. At $T=T_{corr}$ the strong Coulomb correlations within the dimers soften, giving rise to a new monoclinic insulating phase shown in **(c)** via a purely-electronic phase transition. Finally, only after the *e*-*e* correlations are sufficiently diminished ($T=T_{corr}$+3 K), the system undergoes a transition to a rutile metallic phase shown in **(d).** **(e)** Temperature- and polarization-dependent PEEM images measured at the photon energy of the leading slope of the O *K* absorption edge (528.7 eV) which is most sensitive to the collapse of the insulating gap in $VO_2$, and provides a clear contrast mechanism (up to 20%) between the insulating and metallic phases (see Fig. S3 in the Supplemental Material [20]). In contrast to the bulk-like $VO_2$ films, our high-quality (near-ideal) ultrathin epitaxial $VO_2$ film grown on $TiO_2$(001) substrate exhibits a homogeneous insulator-metal transition without phase segregation. Darker defects (likely specs of dust) in the left part of the images were used for fiducial alignment and focusing. The overall intensity of the images increases with temperature, which is consistent with the changes in XAS intensity at 528.7 eV as $VO_2$ undergoes the insulator-metal transition.

# Supplemental Material for

# Correlation-driven insulator-metal transition in near-ideal vanadium dioxide films

In this supplemental section we would like to present some additional materials related to characterization of the samples mentioned in the letter, as well as some expanded temperature-dependent XAS and PEEM datasets detailing the results shown in the main text. These provide additional insights into the validity and interpretation of our experimental data. The supplementary materials include temperature-dependent x-ray diffraction measurements ($\theta$ - $2\theta$) of the $TiO_2$ substrate (002) and the $VO_2$ $(002)_R$ film peaks in the metallic and insulating phases, and expanded dataset showing individual spectra of the temperature-dependent evolution of the $d_{\|}$ V-V singlet peak, $d_{\|}$ Peierls peak, as well as the collapse of the upper band gap.

## Temperature-dependent x-ray diffraction (XRD) measurements

High-angular-resolution (<0.01°) temperature-dependent XRD measurements of the $VO_2/TiO_2(001)$ films were carried out in a vacuum of ~1 mTorr, using a Bruker D8 Discover system equipped with variable-temperature stage enclosed in an x-ray transparent beryllium dome. Figure S1 shows two typical $\theta$-$2\theta$ scans across the $TiO_2$ substrate (002) and the $VO_2$ film peaks, which are indexed as $(002)_R$ for the high-temperature rutile phase, and $(\bar{4}02)_M$ for the low-temperature monoclinic phase. The scans were collected immediately below and above the insulator-metal transition, and show an abrupt change of the out-of-plane inter-planar spacing in the $VO_2$ film at the transition temperature $T_{IMT}$=295K (green circles in the inset). Concomitant in-situ electronic transport measurements (purple circles in the inset) confirm that the insulator-metal transition and the structural monoclinic-rutile transition occur at the same temperature.

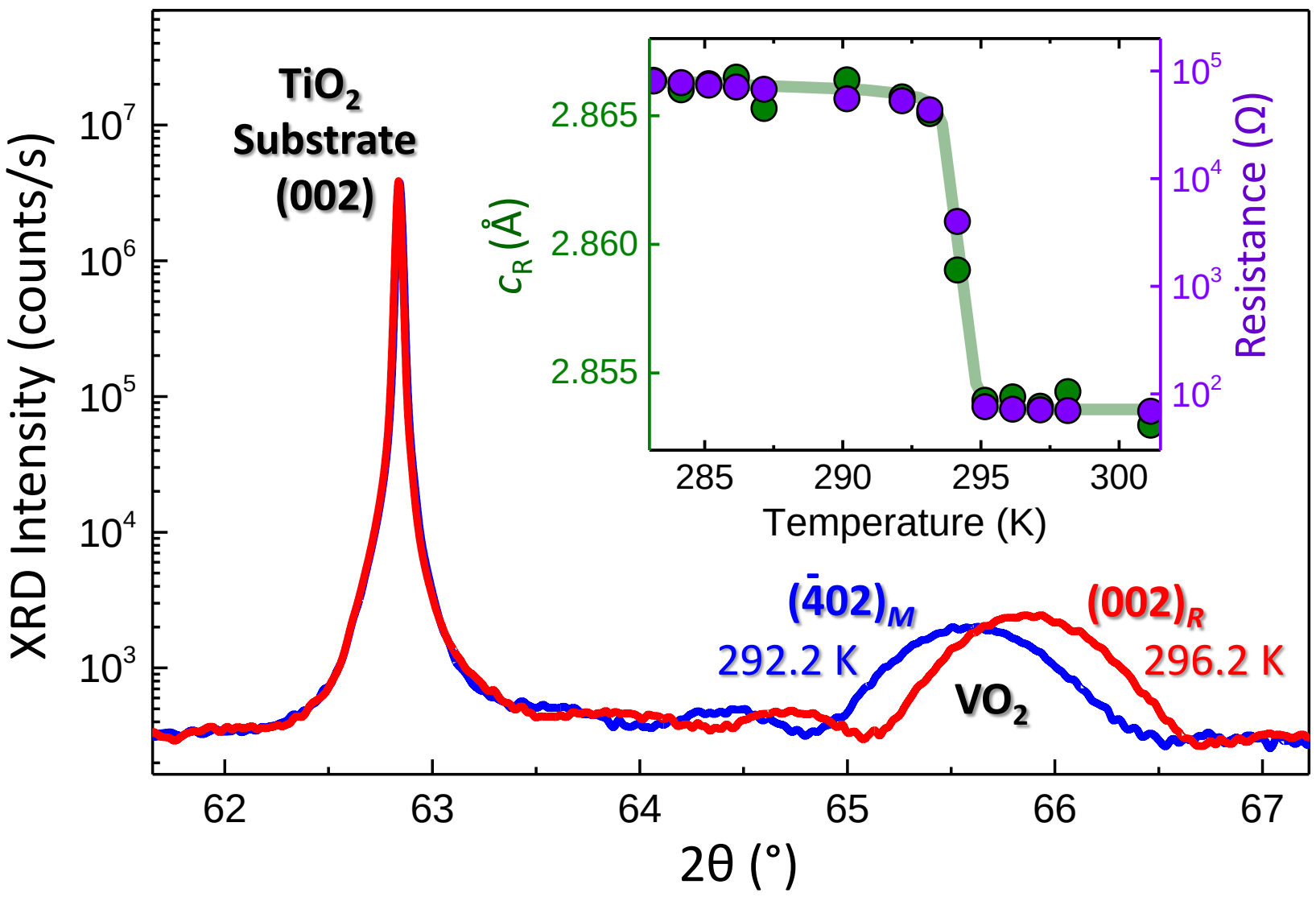

**Figure S1 | Monoclinic/Rutile structural transition in near-ideal coherently-strained $VO_2/TiO_2(001)$.** Temperature-dependent x-ray diffraction $\theta$-$2\theta$ scans across the $TiO_2$ substrate (002) and the $VO_2$ film peaks show a clear structural transition in $VO_2$ manifested by an abrupt change in the inter-planar atomic spacing along the direction normal to the film surface at 295K (inset). Concomitant XRD and electronic transport measurements were collected for the heating cycle, with 1 hour temperature equilibration time between scans. **Inset:** Temperature-dependent resistance (purple circles) and $c_R$ lattice parameter (green circles) obtained concomitantly via electronic transport and x-ray-diffraction measurement respectively.

## Temperature- and polarization-dependent x-ray absorption (XAS) measurements – Expanded Dataset

In this supplementary section we expand on the data shown in Fig. 3a of the main text by showing individual spectra of the temperature-dependent evolution of the $d_{\parallel}$ V-V singlet peak, $d_{\parallel}$ Peierls peak, as well as the collapse of the upper band gap. These provide additional insight into the validity of our experimental data. Since all three datasets are extracted from the same XAS spectra, no temperature calibration is required to compare these two plots. The plots clearly illustrate that the two electronic-structure transitions happen at two distinct temperatures, with $T_{corr}$=290K and $T_{IMT}$=295K.

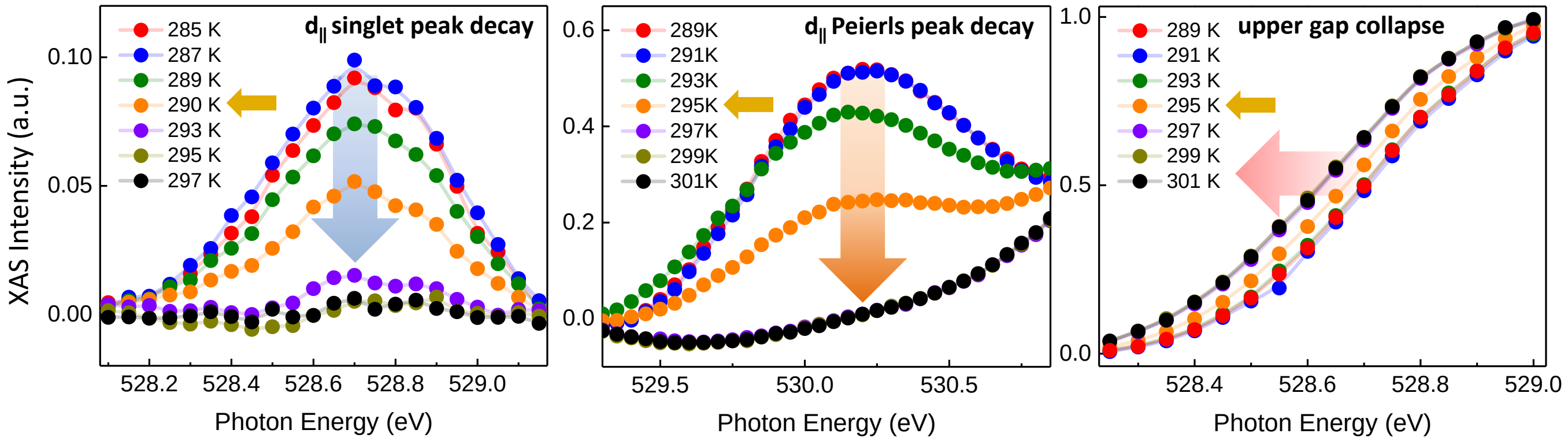

**Figure S2| Two electronic transitions in $VO_2$ – expanded datasets.** Temperature-dependent evolution of the $d_{\parallel}$ V-V singlet peak (left panel) showing distinctly different transition temperature ($T_{corr}$=290 K) as compared to the $T_{IMT}$ (295 K), the critical temperature at which the decay of the $d_{\parallel}$ Peierls peak (middle panel) as well as the collapse of the upper band gap (right panel) is observed in $VO_2$.

## Temperature-dependent PEEM measurements of bulk-like $VO_2$ film – Phase Separation

In this supplementary section we show results of the temperature-dependent photoelectron microscopy (PEEM) measurements of bulk-like $VO_2$ film on $Al_2O_3(10\bar{1}0)$ substrate showing clear separation into metallic and insulating regions during the insulator-metal transition (at 340K), which is in stark contrast with the high-quality ultrathin epitaxial $VO_2$ films grown on $TiO_2(001)$ substrate (see Fig. 3 of the main text). Measurements were carried out at the leading slope of the O K absorption edge (528.8 eV) which is most sensitive to the collapse of the insulating gap. As expected from the XAS spectroscopic characterization of the same sample (plot on the right), metallic phase appears in the form of lighter patches at 528.8 eV.

.

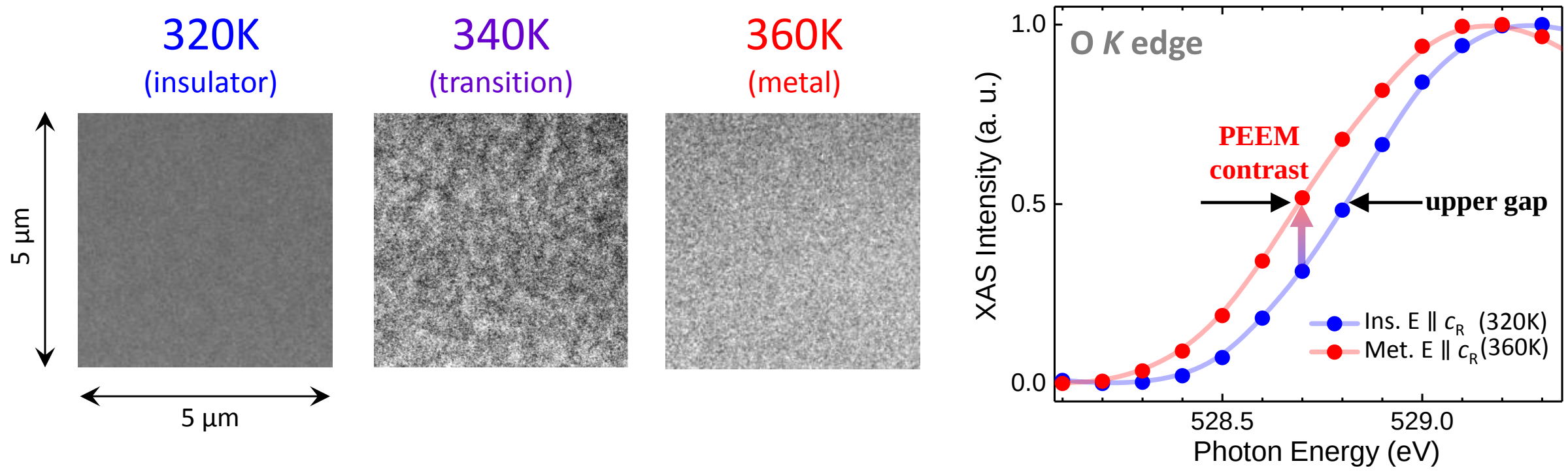

**Figure S3| Phase separation in bulk-like $VO_2$ film.** Temperature-dependent PEEM images of bulk-like $VO_2$ film on $Al_2O_3(10\bar{1}0)$ substrate measured in the insulating state (left panel), during the insulator transition (middle panel) and in the high-temperature metallic state (right panel). Images show clear separation into metallic and insulating regions during the insulator-metal transition (at 340K), in contrast with the high-quality ultrathin epitaxial $VO_2$ films grown on $TiO_2(001)$ substrate described in the main text (see Fig. 3).